\documentclass[prb, preprint, superscriptaddress]{revtex4-1} 

\usepackage{amsmath}  
\usepackage{amsfonts} 
\usepackage{graphicx} 
\usepackage{epstopdf}

\begin{document}

\title{Visible optical beats at the hertz level}

\author{Mickey McDonald}
\email{mpm2153@columbia.edu}
\affiliation{Department of Physics, Columbia University, New York, NY 10027}
\author{Jiyoun Ha}
\affiliation{Tenafly High School, Tenafly, NJ 07670}
\author{Bart H. McGuyer}
\affiliation{Department of Physics, Columbia University, New York, NY 10027}
\author{Tanya Zelevinsky}
\affiliation{Department of Physics, Columbia University, New York, NY 10027}

\date{\today}

\begin{abstract}
We present a lecture demonstration which produces a visible, beating interference pattern that is the optical analog of demonstrations which produce audible, beating sound-wave interference. 
The setup is a compact, portable Mach-Zehnder interferometer made of optical components commonly found in laser physics labs.  
This apparatus may also be built and used in advanced laboratory courses to illustrate concepts in interferometry and laser light modulation.
\end{abstract}

\maketitle 

When presenting demonstrations of the wave nature of light, instructors are often limited to static interference patterns formed by light waves diffracting through stationary slits or passing through an interferometer.
Such static patterns arise from the relative phase differences of multiple wave paths.  
In contrast, to demonstrate the wave nature of sound, instructors most often present dynamic interference patterns using two tuning forks or sine-wave-driven speakers.\cite{Boucher}
Here, a controlled relative frequency difference produces an audible, beating sound envelope.  
While optical analogs of such audio demonstrations exist,\cite{Kawalec, Razdan, Basano1997, Basano2000, Louradour, Jacobs, Corey} they seem to be rarely used, perhaps because of the perceived complexity and cost, and usually rely on electronic detection of the interference.  
In this Note, we present a striking demonstration of dynamic light interference visible by eye, which can be built simply, compactly, and relatively inexpensively by using (or borrowing) optical components commonly found in laser physics labs.

\begin{figure}[b]
\centering
\includegraphics[width=6in]{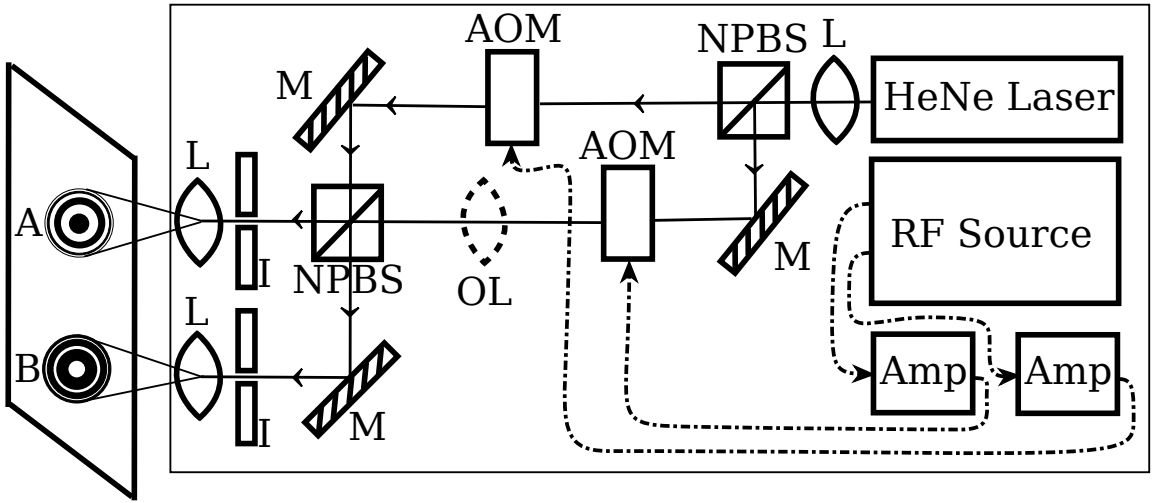}
\caption{Schematic of the demonstration apparatus.  (A video showing the actual apparatus is available as a supplementary media file.)
Laser light following the solid arrows produces two complementary interference patterns, A and B, which visibly beat at an optical frequency difference created by acousto-optic modulators (AOMs).
A lens (L) directly after the laser output focuses light into the AOMs to improve diffraction efficiency, while an optional lens (OL) in one path alters the relative beam curvature, introducing circular fringes.
Two more lenses expand the beams to enhance visibility of the interference pattern.
Irises (I) block unwanted light produced by higher diffraction orders of the AOMs.
Dashed arrows represent radio-frequency (RF) coaxial cables. 
Amp, amplifier; M, mirror; NPBS, nonpolarizing beam splitter; RF Source, two-channel RF frequency synthesizer. 
}
\label{schematic}
\end{figure}

Figure ~\ref{schematic} is a sketch of the apparatus, which is a  Mach-Zehnder interferometer\cite{Zetie} modified to include two acousto-optic modulators (AOMs).
Light from a source laser is split along two optical paths.
Each path contains one AOM, which serves to shift the frequency of the light by an amount equal to the radio frequency (RF) of a sine-wave signal input to each AOM.  This shift originates from the light scattering off traveling pressure waves in a crystal, and resembles the Doppler shift of light scattering off a moving diffraction grating.\cite{Corey}

When the light beams from both paths are recombined, the resulting interference pattern beats at a rate equal to the difference between the optical frequencies of the paths, which is the same as the difference between the frequencies of the RF AOM signals.
This beating of two superimposed beams of equal amplitudes $E_0$ oscillating at optical frequencies $\omega_1$ and $\omega_2$ can be written as
$$|E_0\cos{(\omega_1t)}+E_0\cos{(\omega_2t)}|^2 = 2|E_0|^2[1+\cos{(\omega_bt)}]\cos^2{(\bar{\omega}t)},$$
where the beat frequency $\omega_b = \omega_1-\omega_2$ and the optical carrier frequency $\bar{\omega} = (\omega_1+\omega_2)/2$.
The superposition is squared because our eyes are sensitive to intensity rather than amplitude.
By choosing the RF difference to be 1 Hz, for example, the interferometer output then beats at 1 Hz, which is readily visible by eye. 
RF differences beyond about 20 Hz produce beats that are undetectable by eye, causing the interference pattern to wash out.

\begin{figure}[t]
\centering
\includegraphics[width=6in]{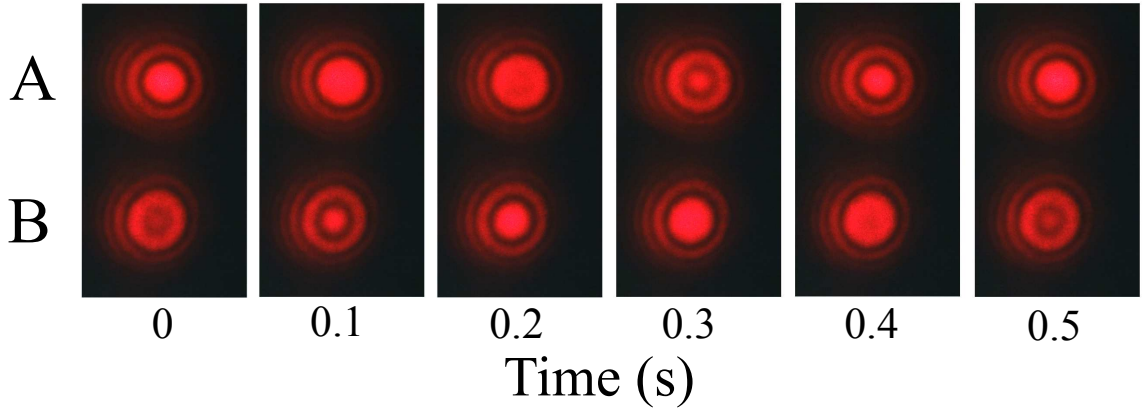}
\caption{(Color online) Images of beating interference patterns A and B created by a 2 Hz optical frequency difference.
A video of this interference is available as a supplementary media file.\cite{YOUTUBE}}
\label{pattern}
\end{figure}

Figure~\ref{pattern} shows a time series of images of 2-Hz dynamic interference patterns produced by our apparatus.  
One benefit of the Mach-Zehnder configuration is that it produces two complementary interference patterns, labeled A and B, which can be viewed together.
These patterns are inverses of one another, as required by energy conservation.  
However, the mechanism behind this inversion is not obvious.
At first glance the beam paths, each involving an equal number of reflections and transmissions, appear to be symmetric.
Careful analysis reveals that a phase shift of either $\pi$ or 0 radians occurs depending on whether light reflects from  a ``hard'' surface ($n_f>n_i$) or a ``soft'' surface ($n_f<n_i$), resulting in complementary patterns. \cite{Zetie}

To enhance visibility, we introduced circular interference fringes with an optional lens (OL in Fig.~\ref{schematic}), which creates different wavefront curvatures for the interfering beams.\cite{Nachman} 
For equal RF AOM frequencies, these circular interference fringes are fixed. 
For unequal RF AOM frequencies, the circular interference fringes move radially inward or outward, depending on the sign of the frequency difference.  
For phase-continuous RF sources, such as the direct-digital-synthesis source we used, the circular fringes move continuously and do not jump, even while the frequency difference is changed.  

There is flexibility in the type of laser used for this demonstration.
In addition to the HeNe laser, we successfully used both red and green laser pointers.
Note that the frequency stability of the source laser does not matter as long as the path-length difference between interferometer paths (typically $\ll$ 1cm) is less than the coherence length of the laser, which usually ranges from 10 cm to hundreds of meters for HeNe lasers.
Inexpensive laser pointers with very short coherence lengths may require more precise interferometer alignment.
The degree of linear polarization of the source light, however, does affect the quality of the patterns A and B.
We found that placing a polarizer on the output of our HeNe laser helped to sharpen the contrast of the interference pattern.

The most expensive components required are the two AOMs and suitable RF frequency sources with Hz resolution, but these can often be found in (or borrowed from) modern undergraduate or research laboratories that use lasers, or purchased inexpensively from surplus sources.  
For our apparatus, we used two identical 80-MHz Isomet model 1205C-2 AOMs (\$800 each).  
However, AOMs with different operating frequencies may be used.  
For example, our initial apparatus used one 80-MHz AOM operated near 95 MHz and another 210-MHz AOM operated near 190 MHz.  
To correct for the frequency difference, we passed light twice through the 80-MHz AOM.
Where acousto-optic modulators are not available, another option for shifting the frequency of laser light is to pass the beam through a transmission diffraction grating moving transverse to the beam.
This technique produces a small Doppler shift $\Delta \nu = \pm mV/d$ in the $m$-th diffracted order of the beam, where $V$ is the velocity of the grating and $d$ is the grating spacing.\cite{Corey}

When using AOMs to provide the beat note, it is critical to drive them with a pair of RF sources capable of providing a stable Hz-level beat frequency.
If independent RF sources are used, their stabilities must be better than $\sim$10$^{-8}$ in order to ensure a steady $\sim$1 Hz difference between AOMs driven at $\sim$100 MHz.
An easier (and more economical) solution is to use a two-channel frequency synthesizer with at least 1-Hz resolution, which will guarantee phase coherence between the two AOM RF inputs.
We used a Novatech model 409B/02 dual-channel RF synthesizer (\$895)
with 1 W amplifiers (RF Bay MPA-10-40 (\$229) and Minicircuits ZHL-3A (\$229)).  
More economical synthesizers such as the Novatech DDS9m/02 (\$645) would be suitable as well.  

In summary, we present a portable, compact, and dynamic demonstration that uses optical self-heterodyning to allow students to directly observe (by eye) the wave nature of light, making use of an interferometer and acoustic-optic modulators to create a hertz-level frequency difference between lasers beams with absolute frequencies of hundreds of terahertz.
We believe this will be a valuable addition to the repertoire of lecture demonstrations of the wave nature of light, and additionally can serve as an educational platform for advanced students to learn the concepts of modulation and interferometry.

\end{document}